\documentclass[superscriptaddress,floatfix,prl,twocolumn,amsmath,amssymb]{revtex4-2}

\usepackage{graphicx}
\usepackage{dcolumn}
\usepackage{bm}
\usepackage{amsthm}
\usepackage{natbib}
\usepackage{mathtools}
\usepackage{color}
\usepackage{footnote}
\usepackage{bbold}
\usepackage[normalem]{ulem}
\usepackage{placeins}
\usepackage{ragged2e}
\usepackage{xfrac}
\usepackage{setspace}
\usepackage{multirow}
\usepackage{xcolor}
\usepackage{flushend}
\usepackage{hyperref}
\usepackage{cleveref}
\usepackage{microtype}
\usepackage{siunitx}
\usepackage{caption}
\usepackage{subcaption}
\usepackage{braket}
\usepackage{cellspace, makecell} %
\DeclareMathOperator{\tr}{tr}

\newcommand{\ketbra}[2]{\ket{#1}\!\!\bra{#2}}

\newtheorem{theorem}{Theorem}

\begin{document}

\author{Akram Youssry}
\affiliation{Quantum Photonics Laboratory and Centre for Quantum Computation and Communication Technology, RMIT University, Melbourne, VIC 3000, Australia}

\author{Alberto Peruzzo}
\email[]{alberto.peruzzo@rmit.edu.au}
\affiliation{Quantum Photonics Laboratory and Centre for Quantum Computation and Communication Technology, RMIT University, Melbourne, VIC 3000, Australia}
\affiliation{Qubit Pharmaceuticals, Advanced Research Department, Paris, France}

\title{Universal programmable waveguide arrays}

\begin{abstract}
Implementing arbitrary unitary transformations is crucial for applications in quantum computing, signal processing, and machine learning. Unitaries govern quantum state evolution, enabling reversible transformations critical in quantum tasks like cryptography and simulation and playing key roles in classical domains such as dimensionality reduction and signal compression. Integrated optical waveguide arrays have emerged as a promising platform for these transformations, offering scalability for both quantum and classical systems. However, scalable and efficient methods for implementing arbitrary unitaries remain challenging.
Here, we present a theoretical framework for realizing arbitrary unitary matrices through programmable waveguide arrays (PWAs). We provide a mathematical proof demonstrating that cascaded PWAs can implement any unitary matrix within practical constraints, along with a numerical optimization method for customized PWA designs. Our results establish PWAs as a universal and scalable architecture for quantum photonic computing, effectively bridging quantum and classical applications, and positioning PWAs as an enabling technology for advancements in quantum simulation, machine learning, secure communication, and signal processing.
\end{abstract}

\maketitle

\section{Introduction}
Arbitrary unitary transformations are fundamental in quantum information processing and in classical fields like signal processing and machine learning. Unitary matrices describe quantum state evolution and are crucial for quantum computing, cryptography, and simulation \cite{shor1997algorithms, bennett1984quantum, lloyd1996universal, nielsen2000quantum}. In classical domains, they enable tasks such as dimensionality reduction, signal compression, and MIMO communication \cite{oppenheim1999discrete, arjovsky2016urnn, paulraj2003spacetime}. Despite their broad applicability, scalable and efficient implementations of arbitrary unitaries remain a challenge, especially for high-dimensional transformations.

Integrated photonics offers a promising platform due to its scalability and speed \cite{bogaerts2020programmable}. Currently, the most common approach to implementing arbitrary unitary transformations employs a mesh of Mach-Zehnder interferometers. While effective for decomposing unitary matrices, this setup introduces circuit bends that increase physical footprint and induce losses, limiting scalability \cite{Reck_1994, Carolan2015}.

Advances in integrated optical waveguide arrays have expanded the capabilities of integrated photonics, supporting a wide range of applications in both quantum and classical domains. Within quantum information processing, waveguide arrays are used for tasks like boson sampling, demonstrating quantum advantage \cite{aaronson2011computational}, and quantum walks, enabling exploration of interference and entanglement \cite{peruzzo2010quantum}. Additionally, these arrays allow implementation of entangling gates \cite{Lahini_2018, chapman2023quantum}. In classical domains, waveguide arrays enable the fractional Fourier transform, providing flexible control over signal and quantum state transformations \cite{weimann2016frft}. Furthermore, in topological photonics, waveguide arrays support robust light propagation, akin to topological insulators, enhancing stability against certain types of disorder \cite{ozawa2019topological}.

Programmable waveguide arrays (PWAs) represent a major advancement in integrated photonics, allowing dynamically configurable quantum circuits \cite{youssry2023experimental, yang2024pwa}. Although PWAs have demonstrated potential \cite{youssry2020modeling, Saygin_2020, Skryabin2021, yang2024pwa, yang2024pwaquantum, youssry2023experimental}, the extent of their capabilities under realistic conditions—such as limited control and finite size—remains an open question.

In this work, we address this gap by introducing a rigorous theoretical framework for implementing arbitrary unitaries using cascaded PWAs. We present a decomposition algorithm to realize any unitary matrix with PWAs even with practical constraints. Based on Trotterization, we provide mathematical proof of the asymptotic scaling of the error in terms of the number of sections and the dimensionality of the array. We show that the error vanishes asymptotically as the number of sections increases. Next, we introduce a numerical optimization-based design method for real-world applications. The results of the analytical and numerical approaches are consistent, showing the superior performance of cascaded PWAs in terms of achievable gate fidelities compared to single-section PWAs. The limited performance of the single-section architecture is inherent and cannot be improved by adjustments to materials or geometry. On the other hand, for the cascaded architecture, we show that we can achieve arbitrary precision by controlling the number of sections.  

Our findings position PWAs as versatile, universal building blocks for photonic quantum computing and as a bridge between quantum and classical applications of unitaries. This approach offers scalable solutions for quantum simulation, machine learning, secure communication, and digital signal processing, establishing a robust foundation for the future of integrated photonic systems.

\section{Model and notation}
Programmable waveguide arrays can be modeled as $d-$dimensional quantum systems with nearest-neighbor coupling Hamiltonians:
\begin{align}
    H &= \sum_{m=1}^{d} {\beta_m \ketbra{m}{m}} \nonumber \\ 
    &+ \sum_{m=1}^{d-1}{ C_{m,m+1}\left(\ketbra{m}{m+1} + \ketbra{m+1}{m}\right)}.
\end{align}
where $\beta_m$ is the propagation constant of the $m^{\text{th}}$ waveguide, and $C_{m,m+1}$ is the coupling coefficient between the $m^{\text{th}}$ and the $(m+1)^{\text{th}}$ waveguides. The physical parameters are assumed to be programmable by external controls:
\begin{equation}
\begin{aligned}
    \beta_m &= \beta_0 + \Delta\beta_m , \quad 1\le m \le d \\
    C_{m,m+1} &= C_0 + \Delta C_m, \quad 1\le m \le d-1.
\end{aligned}
\end{equation}
 In photonic waveguide arrays, this can be achieved by adding electrodes, to which voltages can be applied \cite{youssry2020modeling, yang2024pwa}. This modulates the refractive index of the waveguides, effectively controlling $\beta_m$ and $C_{m,m+1}$. Moreover, we have the physical constraints \cite{CMT}:
\begin{equation}
\begin{aligned}
    \beta_m &> 0, \quad 1\le m \le d \\
    C_{m,m+1} &> 0 \quad 1\le m \le d-1.
\end{aligned}
\label{equ:constraints}
\end{equation}
The first condition represents the fact that we have forward propagation, and the second one is required to ensure that the waveguides are not ``erased'' (i.e. they still have different refractive index compared to the bulk). The resulting evolution is 
\begin{align}
    U = e^{-i H L},
\end{align}
where $L$ is the length of the chip (i.e. propagation length), the equivalent of evolution time. The Hamiltonian is assumed time-independent, which means the photon propagates under the effect of a constant Hamiltonian. The system in this setting is not entirely controllable, in other words we cannot implement an arbitrary $SU(d)$ operation. For instance, the identity and any phase shift gate (i.e. an arbitrary diagonal unitary) cannot be implemented because of the strict positivity constraint of the coupling coefficient. This issue is not present in other physical systems such as superconducting qubits where a phase shift gate is implemented virtually \cite{Krantz_2019}. 

In this paper, we mathematically prove that by cascading a finite number of sections, an arbitrary unitary can be achieved with any desired precision. This cascading turns the Hamiltonian to be time-dependent, where the photon experiences a different Hamiltonian in each section. The evolution in this case has to be expressed using a time-ordered exponential, but because the Hamiltonian is piecewise constant over the sections, the expression is reduced to 
\begin{align}
    U = \prod_{k=1}^{K}e^{-i H_k L_k},
\end{align}
where $H_k$ and $L_k$ are the Hamiltonian and length of the $k^{\text{th}}$ section, $K$ is the total number of sections, and we define the product to be in reverse-order,
\begin{align}
    \prod_{k=1}^{K}U_k =  U_K \cdots U_3 U_2 U_1.
\end{align}
Thus, we can treat this problem using the language of quantum control. While a standard Lie Algebra analysis of the system, presented in the Supplemental Material Note 1, shows the general existence of a solution, the analysis does not take into consideration the physical constraints in \cref{equ:constraints}. Here, we prove that a solution exists by demonstrating a decomposition that meets these constraints. In what follows, $\|\cdot\|$ denotes the operator norm, $\mathbb{Z}$ is the set of integers, and $\mathbb{Z}^{+}$ is the set of strictly-positive integers. 

\section{Theoretical Results}
\begin{figure}
    \centering
    \includegraphics[scale=0.75]{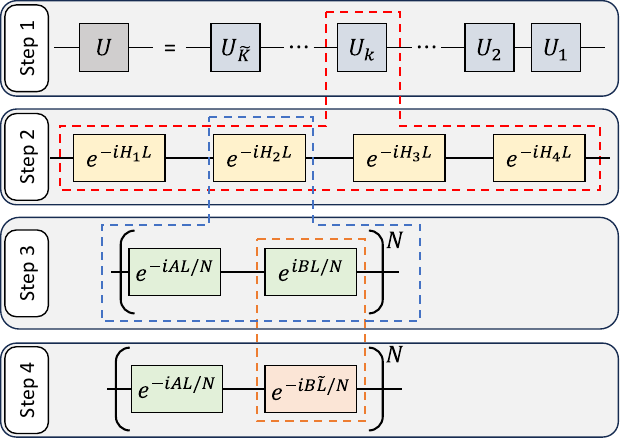}
    \caption{\textbf{The proposed circuit decomposition that satisfies Theorem 1}. Step 1, a general $d \times d$ unitary is decomposed into $\tilde{K}$ 2-mode subspace unitaries acting on  adjacent modes. Step 2, each unitary is decomposed into 4 sections. Step 3, Trotterization is applied to decouple the two modes of interest using $N$ steps. Step 4,  The reverse-time unitary is replaced with a forward-time unitary by altering the section length utilizing the LLL algorithm.}
    \label{fig:Fig1}
\end{figure}

The primary result of this work is captured in the following theorem:
\begin{theorem}
   Let $\tilde{K}=\frac{1}{6}\left(2d^3 - 3 d^2 + d\right)$, and $N\in \mathbb{Z}^{+}$. For $d>2$, an arbitrary $U \in \text{SU}(d)$ matrix, can be decomposed into $K=4\tilde{K} N$ cascaded sections each described by a tridiagonal Hamiltonian with strictly-positive elements. The error between the original matrix $U$ and the decomposition $V$ scales asymptotically as $\|U-V\| = O\left(\frac{\tilde{K}}{N}\right)$. For $d=2$, the decomposition is exact with $K=4$ sections at most. 
\end{theorem}
This theorem establishes the foundation for implementing arbitrary unitary transformations using the PWA architecture. To prove this theorem, we present a 4-step decomposition algorithm (summarized in \Cref{fig:Fig1}) that meets the requirements of the theorem.

\paragraph{Step 1:}
Any matrix $U\in\text{SU}(d)$ can be decomposed into a product two-mode unitaries on a $d-$dimensional space \cite{Reck_1994}. Moreover, we can choose the subspace to be only adjacent-mode by applying additional permutations as needed. The details of such decomposition are shown in Supplementary Note 2. In this case, each section in the cascade acts on two adjacent modes, and leaves the remaining modes unchanged. This requires the $d \times d$ unitary of the $k^{\text{th}}$ section to be in the form:
\begin{align}
    e^{-i H_k L} = I_{m_k} \oplus e^{-i \tilde{H}_k L} \oplus I_{d - m_k -2},
\end{align}
where $H_k$ is the $d \times d$ Hamiltonian of the $k^{\text{th}}$ section, $L$ is the propagation length of one section, $\tilde{H}_k$ is the $2 \times 2$ Hamiltonian acting on the 2-level subspace under consideration, $\oplus$ denotes direct sum operation, $I_a$ is the identity matrix of dimensions $a$, and $m_k$ is an index that determines which two-level subspace we are operating on, which is dependent on $k$. In order to achieve this block structure, the full $d \times d$ Hamiltonian must also have a block-diagonal form:
\begin{align}
    H_k = \left(\frac{2\pi n_1}{L}I_{m_k} \right) \oplus \tilde{H}_k \oplus  \left( \frac{2\pi n_2}{L} I_{d - m_k -2}\right),
\end{align}
where $n_1, n_2 \in \mathbb{Z}$ are arbitrary. This Hamiltonian however is not exactly realizable with the PWA architecture because the waveguides are continuously-coupled, i.e. the strict positivity constraint on the coupling coefficients which prevents a block-diagonal $d \times d$ Hamiltonian. This condition also affects the $2 \times 2$ Hamiltonian preventing achieving an arbitrary $U(2)$ on the subspace. 
For example, implementing a pure $2 \times 2$ phase shift gate requires zero coupling between adjacent waveguides. We shall deal with these two problems next.

\paragraph{Step 2:} To achieve an arbitrary $U(2)$ gates with the chip architecture, we need at most 4 sections as follows. A general $2\times 2$ unitary can be expressed in the form:
\begin{align}
    \tilde{U} = e^{i\eta}\begin{pmatrix}
        r e^{i\phi} & \sqrt{1-r^2} e^{i\delta} \\
       -\sqrt{1-r^2} e^{-i\delta} & r e^{-i\phi}
    \end{pmatrix},
\end{align}
with the parameter $0\le r \le 1$, and the parameters $\eta, \phi, \delta$ can take any value. Let,
\begin{align}
    \xi &= \delta + \frac{\pi}{2}\\
    \zeta &= -(\phi + \xi) \\
    \theta &= \cos^{-1}(r \cos(\zeta)).
\end{align}
The unitary can be decomposed into a cascade of 4 sections with Hamiltonians of the form:
\begin{align}
    \tilde{H}_j = \begin{pmatrix}
        \eta_j + \alpha_j & \kappa_j \\ \kappa_j & \eta_j - \alpha_j
    \end{pmatrix}
    \label{equ:H_j}
\end{align}
where the parameters are given in \cref{tab:Decomp}, and $k_1, k_2, k_3, k_4,  l_2,\in \mathbb{Z}$, are free integers that are chosen according to the physical parameters and their constraints. For example, if one of the values of $\eta_j-\alpha_j$ are negative, the free integer can be chosen to bring back the value to be positive and close to the allowed tuning range. 
\begin{table}[!h]
    \centering \setcellgapes{4pt} \makegapedcells
    \begin{tabular}{|c|c|c|c|c|}
        \hline
         $j$ & 4 & 3 & 2 & 1 \\
         \hline
         $\eta_j$ & $\frac{-\eta + 2\pi k_4}{L}$ & $\frac{-\pi + 4\pi k_3 }{2L}$ &  $\frac{-2\pi k_2}{L} $ & $\frac{-\pi + 4\pi k_1 }{2L}$\\
         \hline
          $\kappa_j$ & $\frac{\sqrt{1-r^2} \theta}{L \sin(\theta)}$ & $\frac{\pi}{2\sqrt{2}L}$ & $\frac{\xi + 2\pi l_2}{L}$ & $\frac{\pi}{2\sqrt{2}L}$ \\
         \hline
         $\alpha_j$ & $\frac{r \sin(\zeta) \kappa_4}{\sqrt{1-r^2}}$ & $\frac{\pi}{2\sqrt{2}L}$ & $0$  & $\frac{\pi}{2\sqrt{2}L}$ \\
         \hline
    \end{tabular}
    \caption{The Hamiltonian parameters of the four sections needed to implement an arbitrary $2\times 2$ unitary gate. The Hamiltonian of each section takes the form \cref{equ:H_j}.}
    \label{tab:Decomp}
\end{table}
The only exception if we are implementing a phase-shift gate $R_z(\xi) = e^{-i \sigma_z \xi}$, which corresponds to $r=0$, then only sections 1, 2, and 3 are needed. The proof of these results is presented in Supplementary Note 3. This step results in exact decomposition of the $2\times 2$ block by increasing the total number of stages by 4 at most. For the following steps, we focus on any of those 4 stages.

\paragraph{Step 3:} We next use an approximation scheme to approximate a block diagonal structure for the $d \times d$ Hamiltonian to arbitrary precision. The idea is to utilize the Lie-Trotter formula for matrix exponentials \cite{trotter1,Suzuki_1976},
\begin{align}
    e^{A+B} = \lim_{N\to\infty}\left(e^{A/N}e^{B/N}\right)^N.
    \label{equ:trotter}
\end{align}
This allows us to approximate the matrix exponential of a sum of two Hamiltonians, via a structure of alternating sections, with the error decreasing as the number of sections increasing. As a result, we can engineer the two alternating sections in a way that effectively removes the coupling between two adjacent waveguides on the unitary level rather than on the physical Hamiltonian level which is unphysical. Thus, the $k^{\text{th}}$ $d \times d$ Hamiltonian $H_k$ can be written as $H_k = A_k - B_k$, allowing us to choose the alternating sections Hamiltonians $A_k$ and $B_k$ with the following parameters:
\begin{align}
    A_k: \quad \quad
   \beta_m &= \begin{cases}
       \tilde{\beta}_0, &m \neq m_k\\
       \eta_k + \alpha_k +\tilde{\beta}_0, &m=m_k\\
       \eta_k - \alpha_k + \tilde{\beta}_0, &m=m_k + 1
   \end{cases} \\
   C_{m,m+1} &= \begin{cases}
       \kappa_k + \tilde{C}_0, &m=m_k \\
       \tilde{C}_0, &m\neq m_k
   \end{cases} \\ \nonumber \\
   B_k: \quad \quad
   \beta_m &= \tilde{\beta}_0 \\ 
   C_{m,m+1} &= \tilde{C}_0
\end{align}
$\tilde{\beta}_0 >0 $ and $\tilde{C}_0>0$ can be chose arbitrarily, while the the $2 \times 2$ parameters $\eta_k, \alpha_k, \kappa_k$ are determined according to the decomposition from Step 2. In practice, we can control the precision of the decoupling, i.e. how close the $H_k$ is to block diagonal form, by controlling the Trotter number $N$, Finally, all the physical parameters need to be strictly positive, so we cannot implement $-B_k$ directly. We address this next.

\begin{figure}
    \centering
    \includegraphics[scale=0.65]{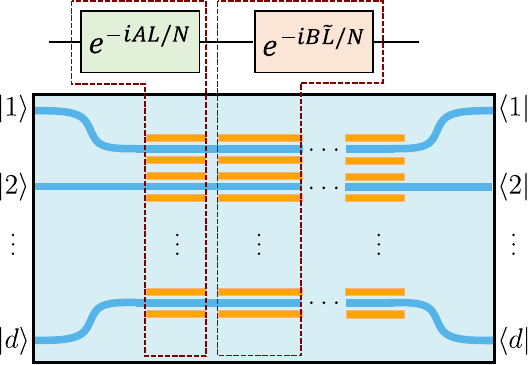}
    \caption{\textbf{Schematic for the architecture of a PWA showing the top view}. The circuit consists of $d$ continuously-coupled waveguides shown in blue, with two surrounding electrodes per waveguide shown in orange.  
    The electrodes are disjoint between sections to allow the independent control of each section. By applying voltage to the electrodes, the propagation constants and the coupling coefficients of the device Hamiltonian can be controlled.}
    \label{fig:2}
\end{figure}

\paragraph{Step 4}: The final step is to apply the quantum recurrence theorem, allowing us to replace the backward time evolution, with forward evolution. This allows us to replace the second alternating section with the Hamiltonian $\-B_k$, with a section of the Hamiltonian $B_k$ but with propagation length (corresponding to the evolution time) $\tilde{L} \gg L$ instead of $L$. The recurrence theorem does not only guarantee the existence of $\tilde{L}$ for $d<\infty$, but also guarantees a controlled precision. Particularly,
\begin{align}
    \forall \epsilon>0, \quad \exists \tilde{L}: \left\|e^{i B_k L} - e^{-i B_k \tilde{L}}\right\| \le \epsilon.
\end{align}
Moreover, this recurrence length can be calculated using a polynomial-time LLL-algorithm \cite{LLL1}, which solves the simultaneous Diophantine approximation problem. The design of section $B_k$ is independent of the particular decomposition, but has a constant tridiagonal form that is also Toeplitz. This allows us to calculate the eigenvalues $\tilde{\lambda}_j$ exactly \cite{tridiagonal} as: 
\begin{align}
    \tilde{\lambda}_j &= \tilde{C}_0 \lambda_j +\tilde{\beta}_0, \quad 1\le j\le d,   \\
    \lambda_j &= -2 \cos{\left(\frac{j \pi}{d+1} \right)} 
\end{align}
Now, applying the LLL-algorithm on the normalized eigenvalues $\lambda_j$, we can find a number $q$ that approximates the eigenvalues to the closest possible integer with any required precision. In this case the evolution is $e^{-i B_k q} \to I_d$. Therefore, if we choose:
\begin{align}
    \tilde{L}= q - \frac{L}{N},
\end{align}
the resulting evolution is $e^{-i B_k \tilde{L}} \to e^{i B_k L/N}$, which is the required form. The precision $\epsilon$ must be chosen so that it does not dominate over the Trotterization error. In Supplementary Note 4, we show that a sufficient choice of the parameters would be:
\begin{align}
    \tilde{C}_0=2\pi j_1, \\
    \tilde{\beta}_0 = \frac{2\pi}{q} j_2,\\
     \epsilon \le \frac{L^2}{2\pi j_1 d N^2},
\end{align}
where $j_1,j_2 \in \mathbb{Z}^{+}$ can be chosen to obtain values close to the tuning range of those physical parameters. We see that all the parameters are independent on the unitary being decomposed, and only depends on the Trotterization number $N$ (which in turns depends on the required overall precision) and the dimensionality $d$. Thus, they can be chosen beforehand during the design process, and does not need to be part of the compilation procedure. The first section $A_k$ will be the only dependent on the unitary and that section is less complex in terms of design. The architecture of the overall scheme is depicted in \Cref{fig:2}.

A final note here, is that practically a small gap is needed to separate the electrodes of two successive sections to avoid short-circuiting. However, the waveguides are still continuously-coupled inside this gap. As a result, the photons experience a zero-voltage Hamiltonian of the form: 
\begin{align}
    H_0 = \beta_0 I_d + C_0\sum_{k=1}^{d-1}\ketbra{k}{k+1} + \ketbra{k+1}{k}
\end{align}
It is clear that the $[H_0, B_k]=0$ for all $k$. This allows us to reconfigure the $B_k$ sections to counter the effect of the free evolution of the photon in the gap. Assuming we have a gap of length $\delta L$ before the electrodes of the $B_k$ sections, a length of $L'$ for those electrodes, and finally another $\delta L$ gap afterwards, then by choosing:
\begin{align}
    \beta' &= \frac{1}{L'} \left(\tilde{\beta_0} \tilde{L} - 2\beta_0 \delta L \right) \\
    C' &= \frac{1}{L'} \left(\tilde{C}_0 \tilde{L} - 2C_0\delta L\right),
\end{align}
we have a new Hamiltonian:
\begin{align}
    B_k' = \beta' I_d + C'\sum_{k=1}^{d-1}\ketbra{k}{k+1} + \ketbra{k+1}{k}.
\end{align}
The propagation through the two gaps and the modified section results then in the desired  unitary,
\begin{align}
    e^{-i H_0 \delta L} e^{-i B_k' L'} e^{-i H_0 \delta L} = e^{-i B_k \tilde{L}}.
\end{align}
\paragraph{Error Analysis:} \cref{equ:trotter} can also be expressed as:
\begin{align}
    &\forall \epsilon_k>0, \quad \exists N: \nonumber \\
    &\left\|e^{-i (A+B) L} - \left( e^{-i A L/N}e^{-i B L/N} \right)^N \right\| \le \epsilon_k.
\end{align}
Moreover, we know that the error scales as \cite{trotter1}:
\begin{align}
    \epsilon_k = O\left(\frac{L^2}{N}\right).
\end{align}
Adding the error of all sections (see Supplementary Note 4), we have the total error:
\begin{align}
    \epsilon_T = O\left(4\tilde{K} \frac{L^2}{N}\right),
\end{align}
where $\tilde{K}=\frac{1}{6}d(d-1)(2d-1)$ is the total number of unitaries in the decomposition due to Steps 1, and the factor of 4 is due to Step 2. The propagation length $L$ of section A is fixed, so the main factor affecting the error is dimensionality $d$ and the Trotter number $N$. This completes the decomposition scheme proving Theorem 1. $\qed$
%%%%%%%%%%%%%%%%%%%%%%%%%%%%%%%%%%
\section{Numerical Results}
Engineering a device based on Theorem 1’s specifications--such as the number of sections determined by the Trotterization error and section lengths set by the recurrence theorem--can be challenging in practice. 
Here we show a numerical study, where we approximately decompose the desired $SU(d)$ unitary, and constraint the number of sections as well as the length of one section of the chip. For the Hamiltonian parameters, we use the model of the proton-exchanged lithium niobate photonic platform as reported in \cite{yang2024pwa}. Electro-optic materials like this are preferable for implementing this architecture due to their ability to minimize cross-talk. 
We assume that,
\begin{align}
    \beta_m &= \frac{2\pi}{\lambda}n_0\left(1 + \frac{\Delta n}{n_0}\Delta V_m \right) \\
    C_{m,m+1} &= C_0\left(1 + \frac{\Delta C \Delta V_{m,m+1}}{C_0}\right)
\end{align}
where $\lambda=808\times10^{-9} \text{m}$ is the wavelength of the photons, $n_0=2.713$ is the refractive index of the waveguide at zero voltage applied to the electrodes, $\Delta n=5\times 10^{-6} \text{V}^{-1}$ is the sensitivity of the propagation constant, $|\Delta V_m| \le V_{\max}$ and $|\Delta V_{m,m+1}| \le V_{\max}$ are the voltages controlling $\beta_m$ and $C_{m,m+1}$, which are dependent on the voltages applied to the electrodes, with $V_{\max}=15 \text{V}$, $C_0 = 100 \text{m}^{-1}$ is the coupling constant at zero voltage, and $\Delta C = 1.4 \text{m}^{-1}\text{V}^{-1} $ is the sensitivity of the propagation constant. We take the length of one section to be $L=6 \times 10^{-3} \text{m}$, and the gap between the sections to be $\delta L=0.1L$. Using gradient-based optimization, we obtain the voltages applied in each section to implement the desired unitary. The objective is to minimize the infidelity $1-F(U,U_T)$ between the target unitary $U_T$, and the overall chip unitary $U$ where,
\begin{align}
    F(U,U_T) = \frac{\left|\tr{\left(U^{\dagger}U_T\right)}\right|^2}{d^2},
\end{align}
To avoid falling in local minima, we run each optimization 48 times and select the best solution (i.e. the one with the lowest infidelity). 

\begin{figure}
    \centering
    \includegraphics[scale=0.65]{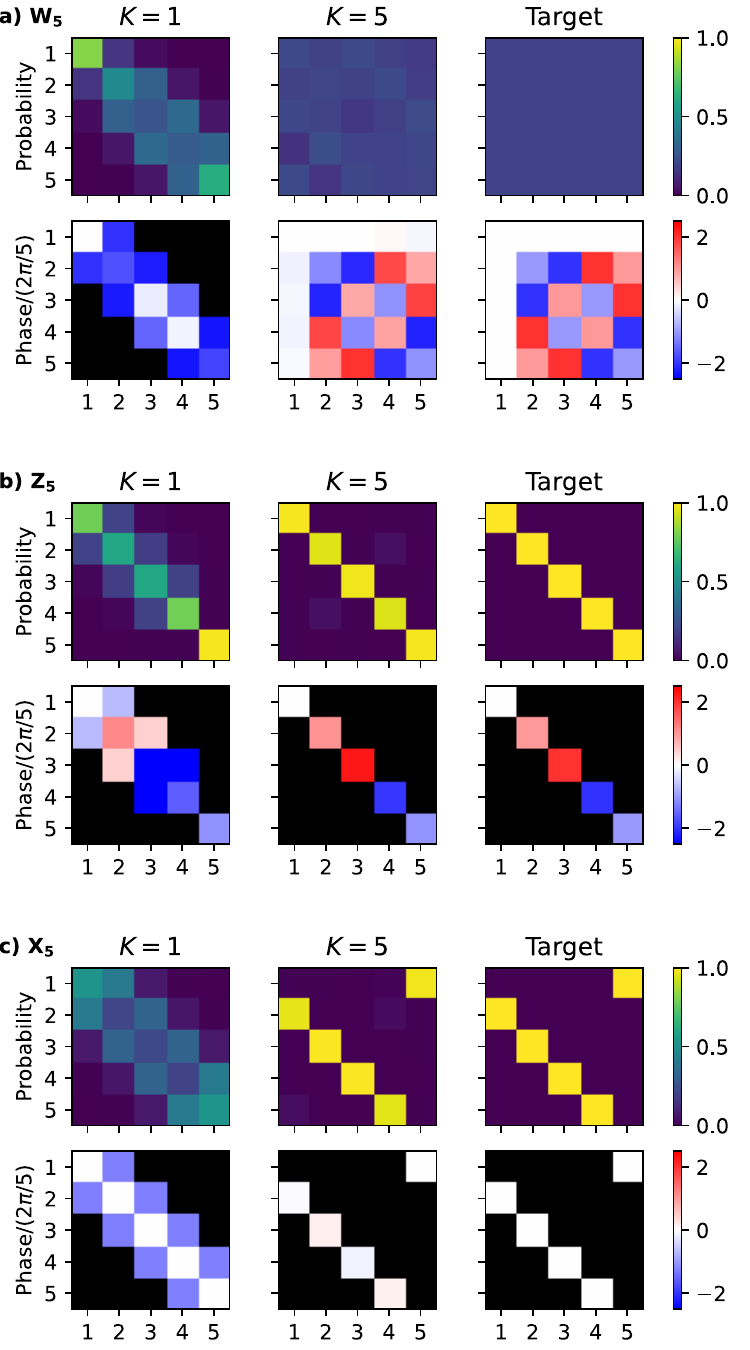}
    \caption{\textbf{Unitary decomposition examples for $d=5$ using numerical optimization.} The plot shows the squared amplitudes and phases of the matrix elements of the optimized unitary for $K=1$ and $K=5$ sections versus the desired target for a) the DFT matrix, b) the clock matrix, and c) the shift matrix. The black color in the the phase shift plots indicate undefined phase because the amplitude is close to zero. The plots have been adjusted to remove the global phase shift. }
    \label{fig:3}
\end{figure}

\begin{figure*}
    \centering
    \includegraphics[scale=0.7]{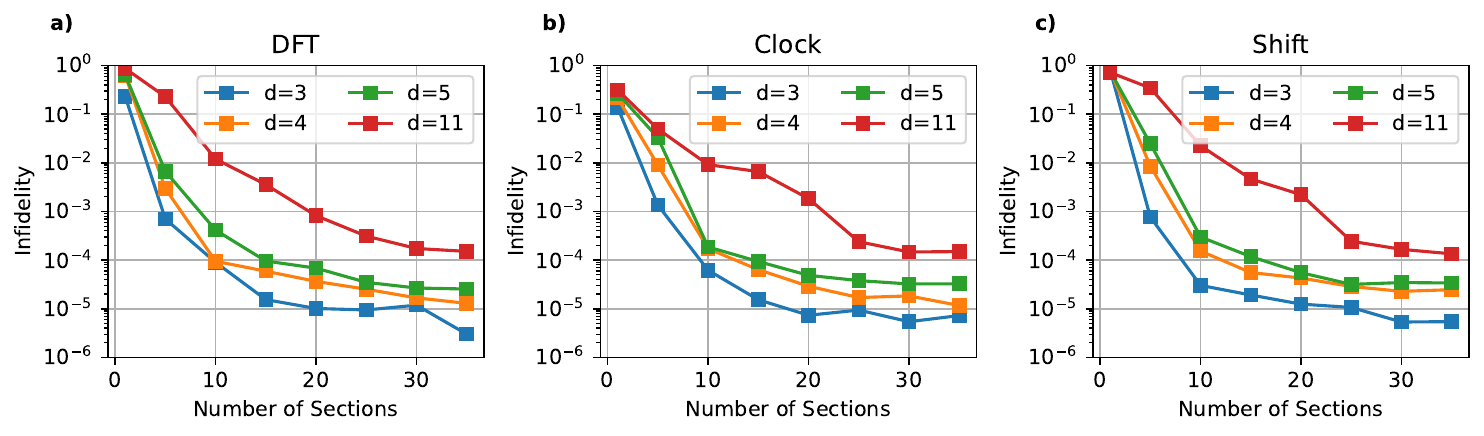}
    \caption{\textbf{The effect of the number of sections and the dimensionality on the infidelity} for a set of 3 numerically-optimized gates, with a section length $L=6\times 10^{-3} \text{m}$. The infidelity is plotted versus the number sections for different dimensions to implement the a) DFT, b) clock, and c) shift unitaries. }
    \label{fig:4}
\end{figure*}

We conduct two sets of numerical experiments. First we study a particular gate set consisting of the $d-$dimensional Discrete Fourier Transform (DFT) matrix $W_d$, the clock matrix $Z_d$, and the shift matrix $X_d$. In computational basis $\{\ket{0}, \ket{1}, \cdots \ket{d-1}\}$, the matrices take the form:
\begin{align}
    W_d &= \frac{1}{\sqrt{d}} \sum_{j,k=0}^{d-1}\omega^{(d-j) k}\ketbra{j}{k} \\
    Z_d &= \sum_{k=0}^{d-1}\omega^{k}\ketbra{k}{k}\\
    X_d &= \sum_{k=0}^{d-1}\ketbra{k+1 \mod d}{k},
\end{align}
where $\omega = e^{2\pi i/d}$. These matrices are important in the study of qudits, and qudit quantum information processing. The clock matrix is a pure phase shift gate, while the shift matrix is a modular permutation gate, transforming each basis state into the next one, with the last state going back to the first state. These gates can be challenging because of the continuous coupling of the waveguides, and the requirement of particular phase shifts. Nonetheless, the numerical results show the possibility of obtaining such gates with high-fidelity for different dimensions. \Cref{fig:3} depicts the optimized unitary the 5-dimensional DFT matrix $W_5$, clock matrix $Z_5$, and shift matrix $X_5$ versus the target, for 1 and 5 sections. \Cref{fig:4} shows more generally the effect of number of sections and dimensionality on the infidelity for the three transformations. Supplementary Figures 1,2, and 3 show similar plots for the cases of $L=1 \times 10^{-3}$m, $L=3.6 \times 10^{-3}$m, and $L=9 \times 10^{-3}$m. Supplementary Figure 4 shows simulations of the propagation of the photons along the chip  (i.e. the probability amplitudes as a function of propagation distance) for $d=5$, and sweeping over all computational bases as initial states utilizing 25 sections. 

In the second set of experiments, we study the performance of a set of 100 Haar random distributed unitaries. \Cref{fig:5} shows a violin plot depicting the statistics of the infidelity versus the number of sections for different dimensionalities. The plot shows concentrated distributions, indicating the success of the optimization procedure despite the physical constraints. 

\section{Discussion}
We have proposed a theoretical framework for implementing arbitrary unitaries based on cascaded programmable waveguide arrays, and provided a proof for the universality of this architecture under the constraints of a tridiagonal Hamiltonian with strictly positive matrix elements. The proof is based on compiling the desired unitary into 2-mode subspace unitaries. Trotterization technique combined with utilization of the LLL-algorithm allow for the decoupling of the modes on the abstract unitary level, rather than on the physical Hamiltonian level which is not allowed. The methods used for the proof can also be applied to higher order couplings, beyond tridiagonal Hamiltonian structure.

Previous studies on universal waveguide arrays, such as in \cite{Saygin_2020} and \cite{Skryabin2021}, lack a rigorous proof of universality. As a result, it is difficult to determine whether the failure to implement a unitary transformation is due to the device’s inherent limitations or simply the result of unsuccessful optimization. Compared to the standard approach for implementing arbitrary unitary operations using a Mach-Zehnder interferometer mesh \cite{Reck_1994}, our architecture, based on an ``always-on'' Hamiltonian, requires only bending at the input and output of the circuit, significantly reducing both the overall footprint and bend-related losses. These bending costs also impact a recent proposal partially based on cascaded waveguide arrays \cite{Saygin_2020}.

\begin{figure}
    \centering
    \includegraphics[scale=0.65]{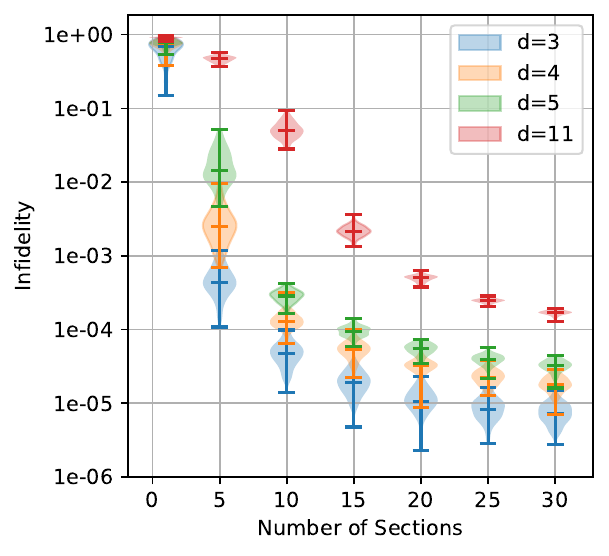}
    \caption{\textbf{The effect of the number of sections and dimensionality on the statistics of the infidelity for a set of 100 numerically-optimized Haar random unitaries.} The violin plot shows the infidelity distribution versus the number of sections. The horizontal lines denote the minimum, median and maximum. }
    \label{fig:5}
\end{figure}

The numerical optimization results presented in this paper are consistent with the theoretical results regarding the foundations of cascaded PWAs. First, in terms of single-section architecture, which is commonly implemented in current technology, we see that it has limited performance. The left column of \Cref{fig:3}, corresponding to $K=1$, showed poor performance for the three gates. There is a significant discrepancy between the target and the optimized unitary in both amplitude and phase shift. This is also seen in \Cref{fig:4}, where the worst-case fidelity, 22.8\%, occurred with the shift gate. Additionally, with one section, it is difficult to overcome the tridiagonal structure in the resulting unitary. The reason is that the propagation constants are orders of magnitude higher than coupling coefficients for realistic materials ($\beta_m \sim 10^{7} \text{m}^{-1}$ versus $C \sim 10^{2} \text{m}^{-1}$). The perturbation methods such as Dyson series can be used to expand the unitary, showing that to first-order, the unitary is tridiagonal (See Supplementary Note 5 for the proof). While a first-order expansion is a rough approximation, we find that it is consistent with the single-section plots in \Cref{fig:3}. Finally, we see from \Cref{fig:5}, that while some of the random gates might perform relatively well in terms of infidelity, they are in fact outliers. The limited performance of single-section PWAs is attributed to the structure and constraints of the Hamiltonian. Therefore, it cannot be improved by increasing the propagation length or choosing different materials. Second, in terms of the proposed cascaded structure, we see that the fidelity improves significantly compared to the single-section architecture. In \Cref{fig:3}, with just 5 sections, both amplitudes and phase shifts of the optimized gates are close to the target. In that case, the fidelity of the shift increased to 97.5\%, and the resulting unitary is no longer constrained to the tridiagonal structure. Because of the time-ordered evolution due to cascading, more generators become accessible, beyond diagonal and first off-diagonal, which expands the controllable space significantly. Figures \ref{fig:4} and \ref{fig:5} show that with increasing the number of sections, the fidelity improves. This is also consistent with the error bound in Theorem 1, that shows that by increasing the Trotter number $N$, the error decreases. The remarkable observation here, is that the numerical optimization solutions are of high quality (infidelities of in the order of $10^{-5}$), with significantly less number of sections compared to the requirements of Theorem 1 that would scale as $K=O(d^3N)$. Thus, while Theorem 1 provides the foundation for the cascaded PWA architecture, in practice the required number of sections is significantly smaller than the theorem requirements. The performance of the cascaded architecture can be further improved with engineering. For example, we see from Supplementary Figures 1, 2 and 3, that by increasing the propagation length of one section, the fidelity improves for the DFT, clock and shift matrices. Equivalently, increasing the electro-optic sensitivity of the material or the maximum allowed voltage range will have similar effect as increasing the propagation length. This gives room for optimizing a design that is suitable for fabrication. Finally, in terms of the dimensionality, we see from Figures \ref{fig:4} and \ref{fig:5} that for a given number of sections, the fidelity decreases with increasing dimensions. Thus, as dimensionality increases, a greater number of sections is required to achieve the same fidelity. This result aligns with Theorem 1, where the error scales as $O(\tilde{K}/N) \sim O(d^3/N)$. 

The Lie-algebra-based techniques employed here can also be applied to other systems. For example, in many physical quantum computing platforms, the qubits are realized by considering a 2-level subspace of higher-dimensional quantum systems. This requires decoupling that subspace perfectly in order to achieve high-fidelity operations. Designing control pulses that reduces the leakage outside the computational subspace is an important challenge. With the method presented here, we can use similar Trotterization approach to decoupling the subspace effectively over the evolution. While the method will result in a long-pulse sequence that could be a challenge in noisy systems such as superconducting qubits, other systems with longer qubit lifetimes such as ion-traps could benefit from that. This can be further investigated in the future.

An accurate model is needed to determine the device parameters post-fabrication to apply the optimization procedure for finding control voltages in individual sections. This can present challenges in practice, making machine learning methods useful in this context. Supervised machine learning in the form of graybox models \cite{youssry2020characterization, youssry2021noise, youssry2022multi}, for example,  was successfully applied to single-section PWA's in \cite{youssry2020modeling}, and verified experimentally in \cite{ youssry2023experimental}. Moreover, reinforcement learning has also been proposed as an alternative for controlling single-section PWAs \cite{Fouad_2024}, and could be extended to the cascaded setting as well.
Future explorations of numerical optimizations could benefit from incorporating methods from geometric gate synthesis \cite{perrier2020quantum}, which may aid in identifying optimal control strategies.
Alternatively, higher-level abstraction corrections, such as the Solovay-Kitaev algorithm \cite{dawson2006solovay} and its recently proposed inverse-free variant \cite{bouland2021efficient}, which employs clock and shift matrices as fundamental building blocks, can also be considered. 
On the theoretical side, other group-theoretic approaches \cite{sawicki2015universality, Sawicki_2017,Mattioli_2023} as well as controllability analysis approaches, including quantum speed limits \cite{qsl1,qsl2,qsl3}, could be explored for obtaining tighter bounds for Theorem 1. \\

\noindent\textbf{Acknowledgements} The authors thank Yang Yang for discussions on the paper. AP acknowledges an RMIT University Vice-Chancellor’s Senior Research Fellowship and a Google Faculty Research Award. This work was supported by the Australian Government through the Australian Research Council under the Centre of Excellence scheme (No: CE170100012). This research was also undertaken with the assistance of resources from the National Computational Infrastructure (NCI Australia), an NCRIS enabled capability supported by the Australian Government.

\bibliography{references}

\end{document}

\author{Akram Youssry}
\address{Quantum Photonics Laboratory and Centre for Quantum Computation and Communication Technology, RMIT University, Melbourne, VIC 3000, Australia}

\author{Alberto Peruzzo}
\email[]{alberto.peruzzo@rmit.edu.au}
\address{Quantum Photonics Laboratory and Centre for Quantum Computation and Communication Technology, RMIT University, Melbourne, VIC 3000, Australia}
\address{Qubit Pharmaceuticals, Advanced Research Department, Paris, France}

\title{SUPPLEMENTARY MATERIALS for \\ Universal programmable waveguide arrays}
\maketitle

\section{Supplementary Note 1: Lie Algebra Analysis of the unconstrained Hamiltonian}
The Hamiltonian of the PWA can be expressed as 
\begin{align}
    H = \sum_{m=1}^{d} {\beta_m \ketbra{m}{m}} + \sum_{m=1}^{N-1}{ C_{m,m+1}\left(\ketbra{m}{m+1} + \ketbra{m+1}{m}\right)},
\end{align}
With 4 electrode structure, we can independently control each term of this Hamiltonian. Therefore, we have $d+d-1=2d-1$ generators, generating the Lie Algebra  (with the commutator being the Lie bracket operation)
\begin{align}
    \mathfrak{g} =\bigg\langle \left\{-i\ketbra{m}{m}\right\}_{1\le m\le d} \cup \left\{-i\left(\ketbra{m}{m+1}+\ketbra{m+1}{m}\right) \right\}_{1\le m \le d-1} \bigg\rangle _{[\cdot,\cdot]}
\end{align}
The elements of the Lie Algebra satisfy the following properties:
\begin{itemize}
    \item $\forall A \in \mathfrak{g}, \alpha\in\mathbb{R}^{+}: -A, \alpha A \in \mathfrak{g}$ 
    \item $\forall A,B \in \mathfrak{g}: [A,B], A+B \in \mathfrak{g}$
\end{itemize}
Using these properties, the goal is to show that $\mathfrak{g} \supset  \mathfrak{su}(d)$, which implies controllability. This can be shown as follows. Define
\begin{align}
    u_{k,j}&=\ketbra{k}{j} +\ketbra{j}{k}\\ 
    v_{k,j}&=-i\left(\ketbra{k}{j} -\ketbra{j}{k}\right) \\
\end{align}
Take an arbitrary generator $\ketbra{k}{k}$, and calculate its commutator with all other generators. We have
\begin{align}
    \big[-i\ketbra{k}{k}, -i\ketbra{r}{r}\big] &= -\ket{k}\braket{k|r}\bra{r} + \ket{r}\braket{r|k}\bra{k} = 0\\
    \big[-i\ketbra{k}{k}, -i\left(\ketbra{r}{r+1} + \ketbra{r+1}{r}\right)\big] &=   \big[-i\ketbra{k}{k}, -i\ketbra{r}{r+1}\big] + \big[-i\ketbra{k}{k},-i\ketbra{r+1}{r}\big]  \\
    &=-\ket{k}\braket{k|r}\bra{r+1} + \ket{r}\braket{r+1|k}\bra{k} -\ket{k}\braket{k|r+1}\bra{r} + \ket{r+1}\braket{r|k}\bra{k} \nonumber \\
    &= \begin{cases}
    -\ketbra{k}{k+1} + \ketbra{k+1}{k}, & r=k\\
    \ketbra{k-1}{k} - \ketbra{k}{k-1}, & r=k-1, k>1 \\
    0, & \rm{otherwise}
    \end{cases}\nonumber \\
    &=\begin{cases}
    -i\left( -i\ketbra{k}{k+1} + i\ketbra{k+1}{k}\right), & r=k\\
    -i\left(i\ketbra{k-1}{k} -i \ketbra{k}{k-1}\right), & r=k-1, k>1 \\
    0, & \rm{otherwise}
    \end{cases}.
    \label{equ:L1}
\end{align}
So, we have
\begin{align}
    \left\{-i v_{k,k+1}\right\}_{1\le k \le d-1}\subset \mathfrak{g}.
\end{align}
Similarly, using those obtained generators, we have
\begin{align}
    \big[-i\ketbra{k}{k}, -i(-i\left(\ketbra{k}{k+1} + i\ketbra{k+1}{k}\right)\big] &=  -i\big[\ketbra{k}{k}, -\ketbra{k}{k+1} + \ketbra{k+1}{k}\big] \\
    &=-i\left(-\ketbra{k}{k+1} - \ketbra{k+1}{k}\right)
\end{align}
\begin{align}
    \left\{-i u_{k,k+1}\right\}_{1\le k \le d-1}\subset \mathfrak{g}.
\end{align}
Next, 
\begin{align}
     \big[-i u_{k,k+1}, -i u_{r,r+1}\big] &= \big[-i\left(\ketbra{k}{k+1}+\ketbra{k+1}{k}\right), -i\left(\ketbra{r}{r+1} + \ketbra{r+1}{r}\right)\big]\\
     &=\big[-i\ketbra{k}{k+1},  -i\ketbra{r}{r+1}\big] + \big[-i\ketbra{k}{k+1}, -i\ketbra{r+1}{r}\big] \nonumber \\
     &+ \big[-i\ketbra{k+1}{k}, -i\ketbra{r}{r+1} \big] + \big[-i\ketbra{k+1}{k}, -i\ketbra{r+1}{r}\big]\\
     &=-\ket{k}\braket{k+1|r}\bra{r+1} + \ket{r}\braket{r+1|k}\bra{k+1} -\ket{k}\braket{k+1|r+1}\bra{r} + \ket{r+1}\braket{r|k}\bra{k+1} \nonumber \\
     & -\ket{k+1}\braket{k|r}\bra{r+1} + \ket{r}\braket{r+1|k+1}\bra{k} -\ket{k+1}\braket{k|r+1}\bra{r} + \ket{r+1}\braket{r|k+1}\bra{k} \nonumber \\
     &=\begin{cases}
     -\ketbra{r-1}{r}+\ketbra{r}{r-1}, &r=k+1 \\
     -\ketbra{r}{r+2} + \ketbra{r+2}{r}, &r=k-1, k>1 \\
     0, &\rm{otherwise}
     \end{cases}
\end{align}
So, we get the new element $-\ketbra{r}{r+2} + \ketbra{r+2}{r} = -i \left(-i\ketbra{r}{r+2} +i \ketbra{r+2}{r}\right):= -i v_{r,r+2}$, this is valid for $1\le r \le d-3$. So, we can now extend the $v$ set to get
\begin{align}
    \left\{-i v_{k,j}\right\}_{1\le k \le d-1, k+1\le j\le k+2}\subset \mathfrak{g}
\end{align}
Similar to calculation \ref{equ:L1}, we can obtain $u_{k,k+2}$ by commuting $\ketbra{k}{k}$ with $v_{k,k+2}$. So know we also have 
\begin{align}
    \left\{-i u_{k,j}\right\}_{1\le k \le d-1, k+1\le j\le k+2}\subset \mathfrak{g}
\end{align}
We can then repeat the calculations with the new generators obtained, in particular from $\big[-i u_{k,k+1}, -i u_{r,r+2}\big]$ we can get the new element $-i v_{k,k+2}$, from which we also obtain the element $-i u_{k,k+2}$ by commuting with $\ketbra{k}{k}$. By repeating these calculations recursively we find that:
\begin{align}
    \left\{-i v_{k,j}\right\}_{1\le k < j \le d} &\subset \mathfrak{g} \label{equ:S1} \\
     \left\{-i u_{k,j}\right\}_{1\le k < j \le d}  &\subset \mathfrak{g}
    \label{equ:S2}
\end{align}
Finally, if we define
\begin{align}
     w_{l}= \sqrt{\frac{2}{l(l+1)}}\left(\sum_{j=1}^{l}{\ketbra{j}{j}} -l \ketbra{l+1}{l+1}\right), \quad 1 \le l \le d-1
\end{align}
from the linearity we get
\begin{align}
    \left\{-i w_l\right\}_{1\le l \le d-1} &\subset \mathfrak{g} \label{equ:S3},\\
    -i I_{d} :=  \sum_{k=1}^{d} {-i\ketbra{k}{k}} &\in \mathfrak{g}. \label{equ:S4}
\end{align}
We know that the set $\{I_d, u_{k,j}, v_{k,j}, w_{k,j}\}_{1\le k \le j < d}$ are the generalized Pauli matrices \cite{Bertlmann_2008}, and they from a complete orthonormal basis of the Hilbert Space $\mathbb{C}^{d}$. Therefore, we can conclude from Equations \ref{equ:S1}, \ref{equ:S2}, \ref{equ:S3}, \ref{equ:S4} that $\mathfrak{g} \supset \mathfrak{u}(d)$, and therefore the chip is controllable. The control is assumed to be piecewise-constant, which corresponds physically to a finite number of concatenations of sections of the chip with various propagation lengths. This analysis however does not generally hold in the presence of constraints.
%%%%%%%%%%%%%%%%%%%%%%%%%%%%%%%%%%%%%%%%%%%%%%%%%
\section{Supplementary Note 2: Decomposition of Unitaries}
Given a matrix $U\in\text{SU}(d)$, it is always possible \cite{Reck_1994} to find the sequence of 2-level subspace unitaries $\{ T_{p,q} \}$ acting on modes $p$ and $q$ such that 
\begin{align}
    \left(\prod_{q=1}^{d-1} \prod_{p=q+1}^{d} T_{p,q} \right) U = I \implies U = \left(\prod_{q=1}^{d-1} \prod_{p=q+1}^{d} T_{p,q} \right)^{\dagger}.
\end{align}
The unitaries $T_{p,q}$ are chosen so that the $(p,q)$ element iteratively becomes zero upon the performing the matrix multiplication, very similar to Gaussian elimination process. Introducing a new indexing $r$ that uniquely determines $(p,q)$ such that 
\begin{align}
    \prod_{r=1}^{(d^2-d)/2} {T_r U} = I.
\end{align}
Let $U^{(0)} = U$, and $U^{(r)}=T_r U^{(r-1)}$ be the unitary at the $r^{\text{th}}$ iteration with the equivalent index $(p,q)$, and denote the matrix elements $u_{ab} = \bra{a}U^{(r-1)}\ket{b}$, then the matrix elements of the required transformation are
\begin{align}
    (T_r)_{ab} = \begin{cases}
        \frac{1}{\sqrt{|u_{qq}|^2 + |u_{pq}|^2}} u_{qq}^*, & a=p,b=p\\
        \frac{1}{\sqrt{|u_{qq}|^2 + |u_{pq}|^2}} u_{pq}^*, & a=p, b=q\\
        -\frac{1}{\sqrt{|u_{qq}|^2 + |u_{pq}|^2}} u_{pq}, & a=q, b=p \\
        \frac{1}{\sqrt{|u_{qq}|^2 + |u_{pq}|^2}}u_{qq}, & a=q, b=q\\
        1, &a=b \neq p,q\\
        0, &\text{otherwise}
    \end{cases}.
\end{align}
This is the identity matrix with four elements at positions $(p,p)$, $(p,q)$, $(q,p)$, and $(q,q)$ replaced by the inverse elements to zero the $(p,q)$ element of $U^{(r-1)}$. The subspace unitaries $T_{p,q}$ requir interaction between the $p^{\text{th}}$ and $q^{\text{th}}$ modes in the chip. However, it is a requirement to restrict interactions only to neighboring modes. Thus, a series of permutations are needed to achieve this requirement. In this case, we have
\begin{align}
    T_{p,q} &= X_{p,q+1}^{\dagger} U_{q,q+1}^{(p,q)}X_{p,q+1}, \quad p>q+1
\end{align}
where $U_{p,p+1}^{(p,q)}$ is the permutated version of $T_{p,q}$ acting only on adjacent modes $q,q+1$, and $X_{a,b}$ is the permutation matrix between modes $a$ and $b$. The permutation can also be decomposed into series of adjacent modes permutations. For example $X_{25} = X_{23}X_{34}X_{45}$.
%$X_{24} = X_{23}X_{34}$.
This results in the decomposition in 
\begin{align}
    T_{p,q} &= \left(\prod_{k=1}^{p-q} X_{p-k,p-k+1} \right)^{\dagger} U_{q,q+1}^{(p,q)}\left(\prod_{k=1}^{p-q} X_{p-k,p-k+1} \right), \quad p>q+1
\end{align}
where all the transformations are now acting on adjacent modes, and thus can be expressed in the block matrix form
\begin{align}
    U_{q,q+1}^{(p,q)} &= \begin{pmatrix}
        I_{q-1} & 0 & 0 \\
        0 & W_{p,q} & 0 \\
        0 & 0 & I_{d-q-1}
    \end{pmatrix}_{d \times d} \\
    X_{k,k+1} &= \begin{pmatrix}
        I_{k-1} & 0 & 0 \\
        0 & X & 0 \\
        0 & 0 & I_{d-k-1}
    \end{pmatrix}_{d \times d}, 
\end{align}
with $I_a$ being the identity matrix of dimensions $a$, and 
\begin{align}
    W_{p,q} &= \frac{1}{\sqrt{|u_{qq}|^2 + |u_{pq}|^2}} \begin{pmatrix} u_{qq}^*
     & u_{pq}^* \\ -u_{pq} & u_{qq}
\end{pmatrix} \\
    X &= \begin{pmatrix}
    0 & 1 \\ 1 & 0
\end{pmatrix},
\end{align}
at each iteration $r$ with corresponding indices $(p,q)$. Note here that while global phase shifts are generally neglected in quantum computations, we cannot neglect a global phase shift affecting a subspace of modes, as it becomes a relative phase shift with respect to the other remaining modes. Furthermore, the reversing operation at the end requires applying $W_{p,q}^{\dagger}$, rather than $W_{p,q}$.

The total number of sections required including the permutations is
\begin{align}
    \tilde{K} &= \left( \sum_{q=1}^{d-2}\sum_{k=1}^{q} 2k + \sum_{q=1}^{d-1}\sum_{p=q+1}^{d} 1 \right)\\
    &= \frac{1}{6}d(d-1)(2d-1),
\end{align}
with the $k^{\text{th}}$ section has the general form unitary
\begin{align}
    U_k = \begin{pmatrix}
        I_{m_k} & 0 & 0 \\
        0 & e^{-i \tilde{H}_k L} & 0 \\
        0 & 0 & I_{d-m_k-2}
    \end{pmatrix}
\end{align}
where $\tilde{H}_k$ is a $2 \times 2$ Hamiltonian, and $m_k$ is an index, and $L$ is the length of propagation. As can be shown in Supplementary Note 3, we need at most 4 sections to implement an arbitrary SU(2) gate, so combined with the results in this section, the total number of section needed is at most $4\tilde{K}=\frac{2}{3}d(d-1)(2d-1)$.
%%%%%%%%%%%%%%%%%%%%%%%%%%%%%%%%%%%%%%%%%%%%%%%%%%
\section{Supplementary Note 3: Decomposition of SU(2) gates}
A general $2\times 2$ unitary can be represented as
\begin{align}
    U &= e^{i\eta}\begin{pmatrix}
        r e^{i\phi} & \sqrt{1-r^2} e^{i\delta} \\
       -\sqrt{1-r^2} e^{-i\delta} & r e^{-i\phi}
    \end{pmatrix}\\
    &= e^{i\eta}\begin{pmatrix}
        r e^{-i\zeta} & \sqrt{1-r^2} e^{-i\frac{\pi}{2}} \\
       -\sqrt{1-r^2} e^{i\frac{\pi}{2}} & r e^{i\zeta}
    \end{pmatrix}\begin{pmatrix} e^{-i\xi} & 0 \\ 0 & e^{i \xi}\end{pmatrix}\\
    &:= e^{i\eta}R\left(r,\zeta, \frac{\pi}{2}\right)R_Z(\xi)
\end{align}
where $\phi = -\zeta - \xi$, $\delta = -\frac{\pi}{2} + \xi$, $R$ is an SU(2) matrix that can be generated from constrained Hamiltonian of the system, and $R_Z$ is a rotation about Z-axis. Next, we show how to derive the Hamiltonian parameters to implement these two gates. 

\subsection{Rotation Gate}
The Hamiltonian of a single section can be expressed as 
\begin{align}
    H &= \begin{pmatrix} \beta_1 & \kappa \\ \kappa & \beta_2  \end{pmatrix} \\
    &= \frac{\beta_1 + \beta_2}{2} I + \frac{\beta_1 - \beta_2}{2} \sigma_z + \kappa \sigma_x \\
\end{align}
Let $\bar{\beta} = \frac{\beta_1 + \beta_2}{2}$, $\alpha=\frac{\beta_1 - \beta_2}{2}$, $\vec{\sigma} = \begin{pmatrix}
    \sigma_x & \sigma_y & \sigma_z
\end{pmatrix}$, and $\vec{n} := \begin{pmatrix} n_x & n_y & n_z \end{pmatrix} = \begin{pmatrix}
    \frac{\kappa}{\sqrt{\alpha^2 + \kappa^2}} & 0 & \frac{\alpha}{\sqrt{\alpha^2 + \kappa^2}}
\end{pmatrix}$, then we have that
\begin{align} 
    H = \bar{\beta} I + \sqrt{\alpha^2 + \kappa^2}\left(\vec{n}\cdot \vec{\sigma}\right).
\end{align}
Let
\begin{align}
    \theta = \sqrt{\alpha^2 + \kappa^2} L,
\end{align}
then the resulting evolution can be expressed as 
\begin{align}
     e^{-i H L} &= e^{-i\bar{\beta} L} e^{-i \left(\vec{n}\cdot \vec{\sigma}\right) \theta} \\
     &= e^{-i\bar{\beta} L} \left( \cos(\theta) I - i (\vec{n}\cdot \vec{\sigma}) \sin(\theta) \right)\\
    &=e^{-i\bar{\beta} L} \begin{pmatrix}\cos(\theta) - i n_z \sin(\theta) & -i n_x \sin(\theta)  \\ -i n_x \sin(\theta) & \cos(\theta) + i n_z \sin(\theta) \end{pmatrix}\\
    &=\begin{cases}
        e^{-i\bar{\beta} L} \begin{pmatrix}
        r e^{-i\zeta} & \sqrt{1-r^2} e^{-i\frac{\pi}{2}} \\
       -\sqrt{1-r^2} e^{i\frac{\pi}{2}} & r e^{i\zeta}
    \end{pmatrix}, \quad &\sin(\theta)>0 \\ \\
    e^{-i\bar{\beta} L} \begin{pmatrix}
        r e^{-i\zeta} & \sqrt{1-r^2} e^{i\frac{\pi}{2}} \\
       -\sqrt{1-r^2} e^{-i\frac{\pi}{2}} & r e^{i\zeta}
    \end{pmatrix}, \quad &\sin(\theta)<0
    \end{cases}
\end{align}
where we have that $0\le r < 
 1$, in addition to the constraint that $\kappa>0 \implies n_x>0$. The case of $r=1$ is not allowed, as it corresponds to $\kappa=0$. Now, we need to solve for the Hamiltonian parameters $\beta_1,\beta_2, $ and $\kappa$. Equating the real parts of the first element we find that 
\begin{align}
    r \cos(\zeta) = \cos(\theta) \implies \theta = \cos^{-1}(r\cos(\zeta)) \in (0,\pi)
\end{align}
This means that $\sin(\theta)>0$. Equating the imaginary parts we get
\begin{align}
    n_x \sin(\theta) = \sqrt{1-r^2} \implies \boxed{\kappa = \frac{\sqrt{1-r^2} \theta}{L\sin(\theta)}>0}
\end{align}
\begin{align}
    -r\sin(\zeta) = - n_z \sin(\theta) &\implies \alpha = \frac{r\sin(\zeta) \theta}{L\sin(\theta)} \\
    &\implies \boxed{\alpha = \frac{r\sin(\zeta) \kappa}{\sqrt{1-r^2}}}
\end{align}
Finally, equating the global phase shift, we get that
\begin{align}
    \boxed{\bar{\beta} = \frac{-\eta + 2\pi k}{L}, \quad k \in \mathbb{Z}}
\end{align}
Then we have simply $\beta_1 = \bar{\beta} + \alpha$ and $\beta_2 = \bar{\beta} - \alpha$. This gives all the required Hamiltonian parameters in terms of the unitary parameters $r, \zeta, \eta$.

\subsection{Phase-shift gate}
This corresponds to the case of $r=1$ in the previous discussion.  Here we need three sections to implement an arbitrary phase shift gate. The idea is that the Pauli $X$ gate can be eigendecomposed into the product of Hadamard gate $\mathcal{H}$, Pauli $Z$, and another Hadamard,  
\begin{align}
    \sigma_x = \frac{1}{\sqrt{2}}\begin{pmatrix}
        1 & 1\\ 1 & -1
    \end{pmatrix}\begin{pmatrix}
        1 & 0\\ 0 & -1
    \end{pmatrix}\frac{1}{\sqrt{2}}\begin{pmatrix}
        1 & 1\\ 1 & -1
    \end{pmatrix}
\end{align}
Because of this eigendecomposition, we can write down a general rotation-about-X gate as 
\begin{align}
    R_X(\xi) &:= e^{-i \xi \sigma_x}\\
     &= \mathcal{H} e^{-i Z \xi} \mathcal{H} \\
     &= \mathcal{H} \begin{pmatrix}
        e^{-i \xi} & 0\\ 0 & e^{i \xi}
    \end{pmatrix} \mathcal{H}.
\end{align}
Now, if we re-arrange the terms and using ($\mathcal{H}^{-1}=\mathcal{H}$ and $\mathcal{H}=\mathcal{H}^{\dagger}$), we get that
\begin{align}
     R_z(\xi) = \mathcal{H} e^{-i \xi \sigma_x} \mathcal{H}.
\end{align}
Now, the Hadamard gate $\mathcal{H}$ can be implemented using one section of the chip because it can be parameterized as 
\begin{align}
    \eta &= \frac{\pi}{2} \\
    \alpha = \kappa &= \frac{\pi}{2\sqrt{2}L}. 
\end{align}
As for the $R_x(\xi)$ gate, is is also implementable with parameters
\begin{align}
    \eta &= 0 \\
    \alpha &= 0 \\
    \kappa &= \frac{\xi}{L},
\end{align}
This is valid for all $\xi \ne 0$. The special case of $\xi=0$, corresponds to identity gate, which can be implemented as a two section of Hadamard or two section of $X$ gates because ($X^2=H^2 = I$). 
%%%%%%%%%%%%%%%%%%%%%%%%%%%%%%%%%%%%%%%%%%%%%%%%%%
\section{Supplementary Note 4: Error Analysis}
Given a section $H_k$, there are two sources of errors that occur when physically implemented: The trotterization, and the recurrence length for the reverse time propagation. However, both errors are controllable, and vanish in the limit as the trotter number $N$ grows. This can be seen as follows. Using Equation 21 in \cite{trotter1}, we have  
\begin{align}
   \left(e^{-i \frac{L}{N} A } e^{i \frac{L}{N}B}\right)^{N}= e^{\left( -i(A - B)L + \frac{1}{2}\frac{L^2}{N}[A,B] +O\left(\frac{L^3}{N^2}\right)\right)}. \label{equ:trotter}
\end{align}
We see that $N\to \infty$, the left hand side converges to $e^{-i(A-B)L}$, which is the required evolution. The strategy is then to choose the length of the second section using the $LLL$ algorithm such that the additional error induced does not dominate the trotterization error. Here we show how to ensure that. 

Let
\begin{align}
    \mathcal{X} = \sum_{m=1}^d \ketbra{m}{m+1} + \ketbra{m+1}{m}, 
\end{align}
then we have 
\begin{align*}
    B = \tilde{\beta}_0 I_d + \tilde{C}_0 \mathcal{X}.
\end{align*}
Let $\lambda_j$ be the eigenvalues of $\mathcal{X}$, and $\tilde{\lambda}_j$ be the eigenvalues of $B$. The eigendecomposition of the Hamiltonian is $B=QDQ^{\dagger}$, and so $e^{-i B \tilde{L}} = Q e^{-i D \tilde{L}} Q^{\dagger}$. The LLL-algorithm is applied to the eigenvalues $\lambda_j$ to obtain the integers $p_j$ and $q$ such that
\begin{align}
    \Delta_j &:= \lambda_j q - p_j \\
    |\Delta_j|&\le \epsilon \label{equ:E3},
\end{align}
with a desired accuracy $\epsilon$. If we choose the parameters $\tilde{C}_0=2\pi j_1$, $\tilde{\beta}_0 = \frac{2\pi}{q} j_2$, where $j_1, j_2 \in \mathbb{Z}^{+}$, and $\tilde{L}=L-(q/N)$
\begin{align}
    e^{-i \tilde{\lambda}_j \tilde{L} } &= e^{-i (\tilde{C}_0\lambda_j + \tilde{\beta}_0)\left(q-\frac{L}{N}\right) } \\
    &= e^{-i 2 \pi j_1 \lambda_j q}e^{-i 2\pi j_2 } e^{i \tilde{\lambda}_j \frac{L}{N}} \\
    &= e^{-i 2 \pi j_1 (p_j + \Delta_j)}e^{i \tilde{\lambda}_j \frac{L}{N}} \\
    &= e^{-i 2 \pi j_1 \Delta_j}e^{i \tilde{\lambda}_j \frac{L}{N}} \label{equ:E1}\\
    &= \left(1 -i 2\pi j_1 \Delta_j + O(\Delta_j^2)\right)e^{i \tilde{\lambda}_j \frac{L}{N}}
\end{align}
As $\epsilon \to 0$, $e^{-i \tilde{\lambda}_j \tilde{L} } \to e^{i \tilde{\lambda}_j(L/N)}$. Now, let the diagonal matrix
\begin{align}
    \Delta = \sum_{j=1}^{d} {2 \pi j_1 \Delta_j \frac{N}{L}\ketbra{j}{j}},
\end{align}
and define the error matrix
\begin{align}
    E &= Q \Delta Q^{\dagger} \implies [B,E] = 0 \label{equ:E2}\\
    e^{-i E (L/N)} &= Q e^{-i \Delta (L/N)} Q^{\dagger}.
\end{align}
The operator norm of the matrix $E$ can be calculated as follows.
\begin{align}
    \|E\| &= \|  Q \sum_j {2 \pi j_1 \Delta_j \frac{N}{L}\ketbra{j}{j}} Q^{\dagger} \| \\
    &=  \|  \sum_j {2 \pi j_1 \Delta_j \frac{N}{L}\ketbra{j}{j}}  \| \\
    & \le 2 \pi j_1 \frac{N}{L}  \sum_{j=1}^{d} |\Delta_j|  \|  \ketbra{j}{j} \| \\
    &=   2 \pi  j_1 \frac{N}{L}  \sum_{j=1}^{d} |\Delta_j| \\
    &\le  2 \pi j_1 d \frac{N}{L} \epsilon.
\end{align}
The second line follows from the invariance of operator norm under unitary transformation, the third line applies the triangle inequality, the fourth line we have the operator norm of a basis state is exactly equal to 1, and the last line follows from the results of LLL algorithm in \cref{equ:E3}. Now, if we choose
\begin{align}
    \epsilon \le \frac{L^2}{2\pi j_1 d N^2},
\end{align}
then
\begin{align}
    \|E\| \le \frac{L}{N} \implies \lim_{N \to \infty} \|E \| = 0.
\end{align}
Going back to the evolution of section $B$, we find now that
\begin{align}
    e^{-i B \tilde{L}} &=  Q \sum_{j=1}^{d} e^{-i \tilde{\lambda}_j \tilde{L}} \ketbra{j}{j} Q^{\dagger} \\
    &=  Q \sum_{j=1}^{d} e^{-i 2 \pi j_1 \Delta_j}e^{i \tilde{\lambda}_j \frac{L}{N}} \ketbra{j}{j} Q^{\dagger}\\
    &=  Q \sum_{j=1}^{d} e^{-i 2 \pi j_1 \Delta_j} \ketbra{j}{j} Q^{\dagger}   Q \sum_{m=1}^{d} e^{i \tilde{\lambda}_m \frac{L}{N}}  \ketbra{m}{m}Q^{\dagger}\\
    &= e^{-i E (L/N)} e^{i B (L/N)}\\
    &= e^{-i (-B + E) \frac{L}{N}}.
\end{align}
The second line is substitution from \cref{equ:E1}, the third line uses the fact that $QQ^{\dagger}=I_d$ and decomposing the diagonal matrix into a product of two diagonal matrices, the last line uses the commuting property in \cref{equ:E2}. Substituting this in \cref{equ:trotter} we get
\begin{align}
   \left(e^{-i \frac{L}{N} A } e^{-i \tilde{L} B}\right)^{N} &= \left(e^{-i \frac{L}{N} A } e^{-i \frac{L}{N}(-B +E) }\right)^{N} \\
   &=e^{\left( -i(A - B + E)L + \frac{1}{2}\frac{L^2}{N}[A,B+E] +O\left(\frac{L^3}{N^2}\right)\right)}\\
   &=e^{\left( -i(A - B)L + \frac{1}{2}\frac{L^2}{N}[A,B] - i E L  -\frac{1}{2}\frac{L^2}{N}[A,E]   +O\left(\frac{L^3}{N^2}\right)\right)}
\end{align}
Because $\|E L\| = O\left(\frac{L^2}{N}\right)$, we see that all the error terms in the exponential are of that same order. Therefore, the error of the $k^{\text{th}}$ stage is
\begin{align}
    \epsilon_k =  O\left(\frac{L^2}{N}\right).
\end{align}
The total error of the decomposition can be calculated straightforward as follows. Let $U = 
U_{\tilde{K}} \cdots U_2 U_1$ be the exact decomposition, while $V = V_{\tilde{K}} \cdots V_2 V_1$ be the decomposition after applying the trotterization and the recurrence theorem. Then
\begin{align}
    \|U-V\| &= \|(U_{\tilde{K}} \cdots U_2 U_1) - (V_{\tilde{K}} \cdots V_2 V_1)\| \\
    &= \| (U_{\tilde{K}} \cdots U_2 U_1) - (U_{\tilde{K}} \cdots U_2 V_1) + (U_{\tilde{K}} \cdots U_2 V_1) - (V_{\tilde{K}} \cdots V_2 V_1) \| \\
    &\le \| (U_{\tilde{K}} \cdots U_2 U_1) - (U_{\tilde{K}} \cdots U_2 V_1) \| + \| (U_{\tilde{K}} \cdots U_2 V_1) - (V_{\tilde{K}} \cdots V_2 V_1)\| \\
    &= \|(U_{\tilde{K}} \cdots U_2)(U_1-V_1)\| + \|\left((U_{\tilde{K}} \cdots U_2) - (V_{\tilde{K}} \cdots V_2) \right)V_1\| \\
    &= \|U_1 - V_1 \| + \| (U_{\tilde{K}} \cdots U_2) - (V_{\tilde{K}} \cdots V_2) \|
\end{align}
In the third line, we applied the triangle inequality. In the last line, we used the fact that the norm is invariant under unitary transformation. Now, if we repeat the same steps again on the second term recursively, we get
\begin{align}
    \| U-V \| &\le \sum_{k=1}^{\tilde{K}} \|U_k -V_k\| 
    \\ &= O\left(\frac{\tilde{K} L^2}{N} \right).
\end{align}
%
%%%%%%%%%%%%%%%%%%%%%%%%%%%%%%%%%%%%%%%%%%%%%%%%%
\section{Supplementary Note 5: First-order expansion of a single-section Hamiltonian}
The Hamiltonian of the chip can be expressed in general as $H = H_0 + V$, where
\begin{align}
    H_0 &=\sum_{m=1}^{d}\beta_m\ketbra{m}{m} \\
    V &= \sum_{m=1}^{d-1} C_{m,m+1}(\ketbra{m}{m+1} + \ketbra{m+1}{m}) 
\end{align}
We move to the interaction picture with respect to $H_0$. Since $H_0$ is a diagonal matrix, then we have
\begin{align}
    U_0(z) &= e^{-i H_0 z} = \sum_{m=1}^{d}e^{-i \beta_m z} \ketbra{m}{m}.
\end{align}
Then we calculate the interaction Hamiltonian in the interaction picture:
\begin{align}
    V_I(z) &= U_0^{\dagger} V U_0 \\
    &=\left(\sum_{k=1}^{d}e^{i \beta_k z} \ketbra{k}{k}\right)\left(\sum_{r=1}^{d-1} C_{r,r+1}(\ketbra{r}{r+1} + \ketbra{r+1}{r})\right)\left(\sum_{m=1}^{d}e^{-i \beta_m z} \ketbra{m}{m}\right)\\
    &= \left(\sum_{k,r,m}e^{i \beta_k z}e^{-i \beta_m z} C_{r,r+1}\ket{k}\braket{k|r}\braket{r+1|m}\bra{m}\right) + \left(\sum_{k,r,m}e^{i \beta_k z}e^{-i \beta_m z} C_{r,r+1}\ket{k}\braket{k|r+1}\braket{r|m}\bra{m}\right)\\
    &= \sum_{k=1}^{d-1} C_{k,k+1}e^{i (\beta_{k}-\beta_{k+1}) z}\ketbra{k}{k+1} + \sum_{r=1}^{d-1}C_{r,r+1}e^{i (\beta_{r+1}-\beta_r) z}\ketbra{r+1}{r} \\
    &=\sum_{k=1}^{d-1}C_{k,k+1}\left(e^{i (\beta_{k}-\beta_{k+1}) z}\ketbra{k}{k+1} +e^{-i (\beta_{k}-\beta_{k+1}) z}\ketbra{k+1}{k}\right)
\end{align}
The interaction unitary $U_I$ expanded up to first-order is
\begin{align}
    U_1(L) &= I - i\int_{0}^{L}V_I(z) dz\\ 
    &= I - i\sum_{k=1}^{d-1}C_{k,k+1}\left(\int_{0}^{L}e^{i (\beta_{k}-\beta_{k+1}) z}dz \ketbra{k}{k+1} + \int_{0}^{L}e^{-i (\beta_{k}-\beta_{k+1}) z}dz \ketbra{k+1}{k}\right) \\ 
    &=I - i\sum_{k=1}^{d-1}C_{k,k+1}\left(f_k(L)\ketbra{k}{k+1} + f_k^{*}(L)\ketbra{k+1}{k}\right)\\
    f_k(L)&:=\int_{0}^{L}e^{-i (\beta_{k}-\beta_{k+1}) z}dz=\begin{cases}
        L & \beta_k=\beta_{k+1} \\
        \frac{e^{-i(\beta_k - \beta_{k+1})L} - 1}{-i(\beta_k-\beta_{k+1})} & \beta_k \neq \beta_{k+1}
    \end{cases}
\end{align}
Moving back to the Schr\"{o}dinger picture we get the total unitary $U(L)$ expanded up to first-order,
\begin{align}
    U(L) &= U_0(L) U_I(L)\\
    &\approx  \sum_{m=1}^{d}e^{-i \beta_m L} \ketbra{m}{m} + \sum_{m,k}e^{-i \beta_m L} C_{k,k+1}f_k(L) \ket{m}\braket{m|k}\bra{k+1} + \sum_{m,k}e^{-i \beta_m L} C_{k,k+1}f_k^{*}(L) \ket{m}\braket{m|k+1}\bra{k}  \\
    &=\sum_{m=1}^{d}e^{-i \beta_m L} \ketbra{m}{m} + \sum_{k=1}^{d-1}e^{-i \beta_k L} C_{k,k+1}f_k(L)\ketbra{k}{k+1} + \sum_{k=1}^{d-1}e^{-i \beta_{k+1}L} C_{k,k+1}f_k^{*}(L)\ketbra{k+1}{k}.
\end{align}
We see that the up to first-order, the resulting unitary is approximately tridiagonal. While this is a rough approximation, in practice for realistic materials, we have $\beta_m \gg C_{m,m+1}$, thus we have that $\|H_0\| \gg \|V\|$, making perturbative expansion very close to the actual unitary $U=e^{-i(H_0 +V)L}$.
%%%%%%%%%%%%%%%%%%%%%%%%%%%%%%%%%%%%%%%%%%%%%%%%%%
\section{Supplementary Figures}
\begin{figure*}[h]
    \centering
    \includegraphics[scale=0.65]{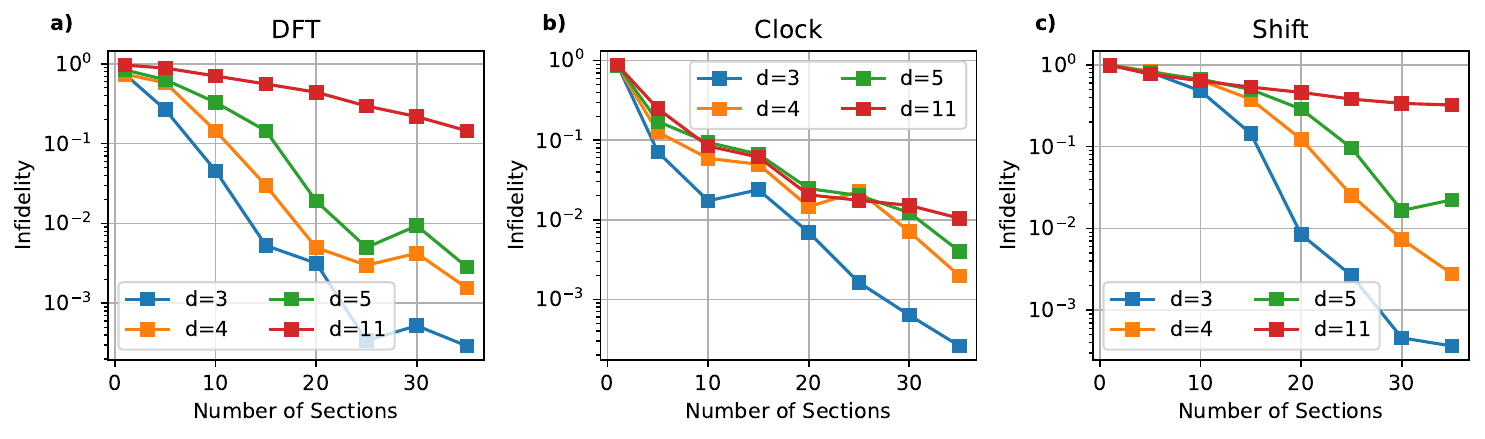}
    \caption{\textbf{The effect of the number of sections and the dimensionality on the infidelity for a set of 3 numerically-optimized gates, with a section length $L=1\times 10^{-3} \text{m}$.} The infidelity is plotted versus the number sections for different dimensions to implement the a) DFT, b) clock, and c) shift unitaries. }
    \label{fig:1mm}
\end{figure*}
\begin{figure*}[h]
    \centering
    \includegraphics[scale=0.65]{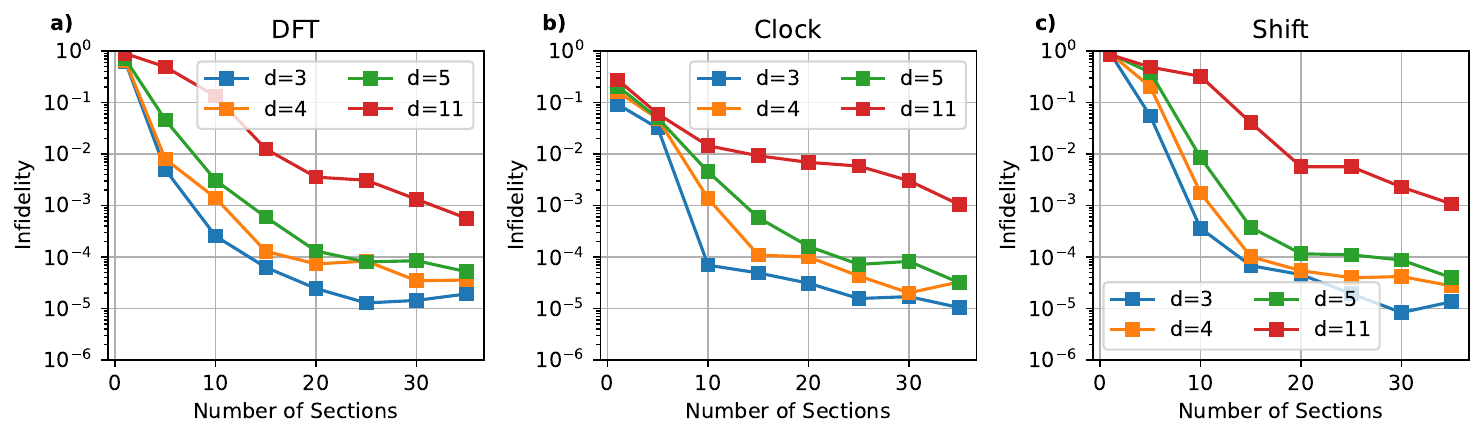}
    \caption{\textbf{The effect of the number of sections and the dimensionality on the infidelity for a set of 3 numerically-optimized gates, with a section length $L=3.6\times 10^{-3} \text{m}$.} The infidelity is plotted versus the number sections for different dimensions to implement the a) DFT, b) clock, and c) shift unitaries. }
    \label{fig:3.6mm}
\end{figure*}
\begin{figure*}[h]
    \centering
    \includegraphics[scale=0.65]{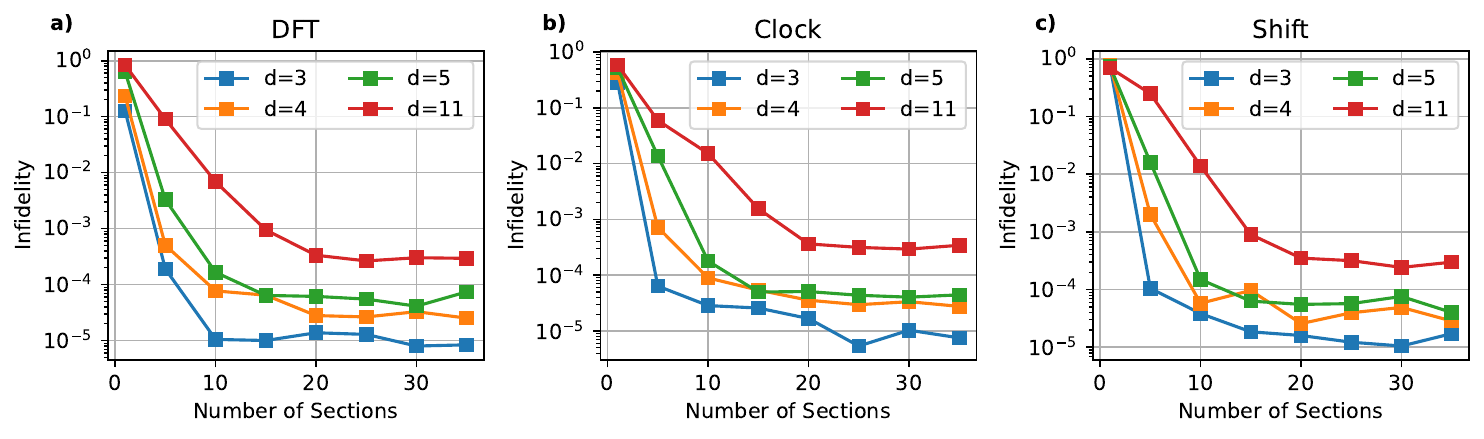}
    \caption{\textbf{The effect of the number of sections and the dimensionality on the infidelity for a set of 3 numerically-optimized gates, with a section length $L=9\times 10^{-3} \text{m}$.} The infidelity is plotted versus the number sections for different dimensions to implement the a) DFT, b) clock, and c) shift unitaries. }
    \label{fig:9mm}
\end{figure*}

\clearpage 
\begin{figure*}[h]
    \centering
    \includegraphics[scale=0.6]{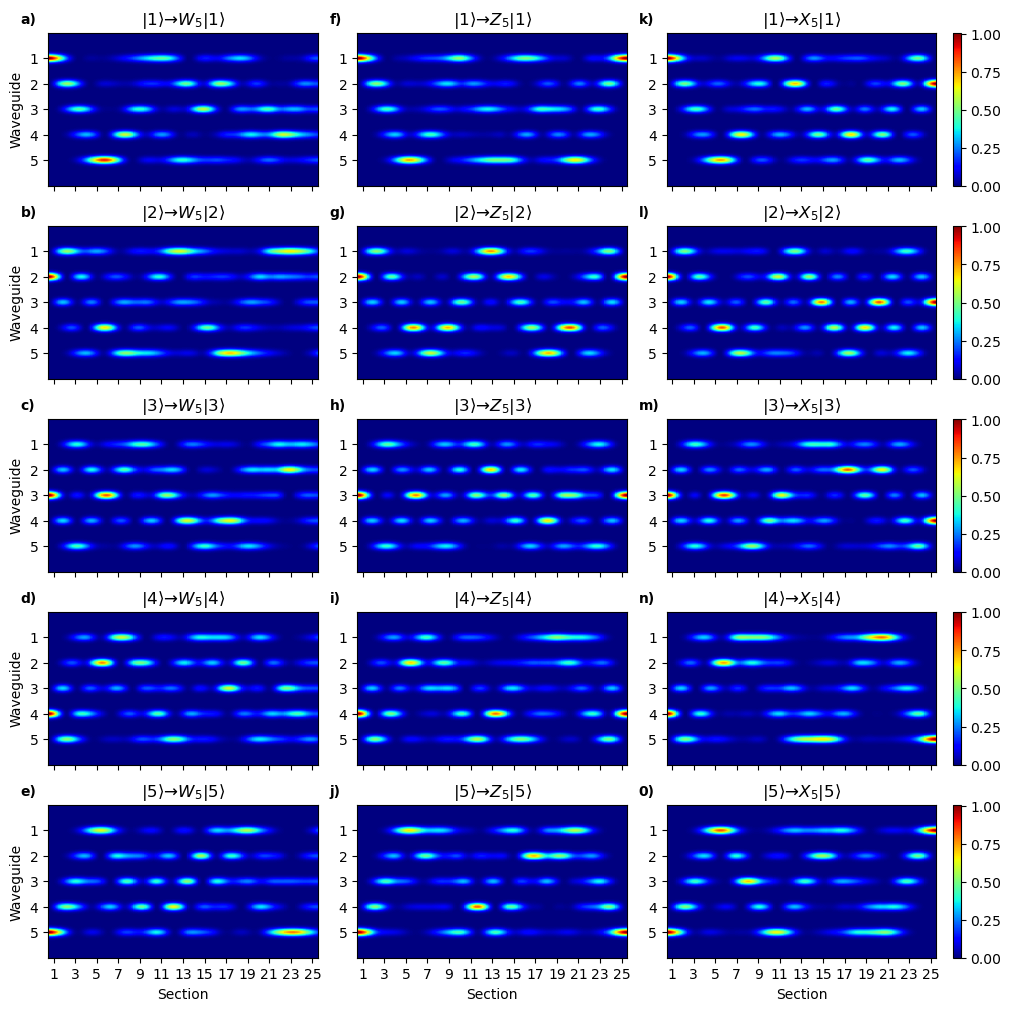}
    \caption{\textbf{Simulation of the mode propagation for a 25-section PWA and section length $L=6 \times 10^{-3} \text{m}$.} The probability distribution of the quantum state along the propagation direction starting from one of the basis states, for a) - e) the DFT gate $W_5$, f) - j) the clock gate $Z_5$, and k) - o) the shift matrix $X_5$.}
    \label{fig:propagation}
\end{figure*}

\bibliography{references}